# Bianchi Type-III Dark Energy Model in f(R,T) gravity with a variable Deceleration parameter.

D.R.K.Reddy[1], G.Ramesh[2], S.Umadevi[2]

[1] AppliedMaths Dept., Andhra University, Visakhapatnam, A.P., India.
[2] Engineering Mathematics, A.U.College of Engineering (Aut), Visakhapatnam, India

**ABSTRACT**:

In this paper, we have investigated Bianchi Type-III Dark Energy Model in f(R,T) theory of gravity proposed by Harko et al. (Phys. Rev. D**84** : 024020, 2011). To find a determinate solution of the field equations we have used (i) the fact that scalar expansion of the space-time is proportional to the shear scalar and (ii) a variable deceleration parameter given by Mishra et al. (Int. J. Theor. Phys. 52: 2546, 2013). Physical and kinematical properties of the model are also discussed. It is observed that our model is in agreement with the recent cosmological observations.

Keywords: f(R,T) gravity, dark energy model. Bianchi-III model. Variable deceleration parameter.

## 1. Introduction

At present we are living in the universe which is expanding and accelerating. This has been established by the Cosmological observations (Perlmutter et al. 1999; Riess et al. 1998; Spergel et al. 2003). Cosmic microwave background (CMB) anisotropy (Caldwell 2002; Huange et al. 2006) and large scale structure strongly indicate that dark energy dominates the present universe, causing the cosmic acceleration. Dark energy and cosmic acceleration are, still, the mysteries of modern cosmology. However, dark energy has, conventionally, been characterized by the equation of state (EoS) parameter

w = $p/\rho$ which is not necessarily constant. Several candidates such as quintessence (Barrien et al. 2000) Phantom (Caldwell et al. 2003) tachyon (Gibbons 2002), k-essence (Chiba et al. 2000), Chapligin gas (Bento et al. 2002) etc. have been proposed to explain the dark energy. However, none of these can be regarded as entirely convincing so far. Another recent approach to explain the late time cosmic acceleration is to modify the general theory of relativity. Among the various modifications, the most important are f(R) gravity, f(T) gravity, f(R,T) gravity etc. In f(R) gravity, gravitational action is modified by adding a function f(R) (R being the Ricci scalar curvature) to Einstein-Hilbert Lagrangian where f(R) provides a gravitational alternative for dark energy causing late time acceleration of the universe. A nice review of f(R) gravity and cosmological models of f(R) gravity is presented by Copeland et al. (2006).

In spite of the above attempts to explain cosmic acceleration, it is still a challenging problem in modern cosmology. Recently, Harko et al. (2011) formulated a modified gravity known as f(R, T) gravity for the said purpose. In this theory, the gravitational Lagrangian is given by an arbitrary function of the Ricci scalar R and of the trace T of the energy tensor. They have obtained field equations in metric formalism and several models in this theory corresponding to some explicit forms of the function f(R,T).

Spatially homogeneous and anisotropic Bianchi cosmological models are important in the discussion of large scale behavior of the universe and such models have been widely studied in general relativity to study the realistic picture of the universe in its early stages. Bianchi type cosmological models in the presence of perfect fluid, in this theory, have been investigated by several authors. In particular, Adhav (2012), Sharif and Zubair (2012) Reddy et al (2012a), Reddy et al (2013), Mishra and Sahoo (2014), Sahoo et al (2014), Sahoo and Mishra (2014) are some of the authors who have discussed Bianchi type models in f(R, T) gravity. Very recently, Sahoo and Sivakumar (2015) obtained LRS Bianchi Type-I cosmological model in f(R, T) gravity with Λ (t) using linearly varying deceleration parameter proposed by Akarsu and Dereli (2012) while Reddy et al (2012b) discussed LRS Bianchi Type-II model with constant deceleration parameter given by Berman (1983). Also Mishra et al (2013) have proposed a linearly varying deceleration parameter and using it they have investigated Bianchi Type-II dark energy model in f(R, T) gravity.

The above discussion and the investigations have inspired us to take up the investigation of dark energy model in Bianchi type-III space time in f(R, T) gravity in the presence of anisotropic dark energy. The present work is organized as follows: Sect.2 deals with a brief review of f(R, T) gravity. In Sect 3, the metric and the explicit field equations of the theory are presented. In Sect.4 exact solutions of the field equations have been given. Sect.5 deals with solutions of the field equations and the model. The physical and kinematical behavior of the model is discussed in Sect.5. The last section contains some conclusions and summary.

## 2. A brief review of f(R, T) gravity

In f(R, T) gravity proposed by Harko et al.(2011), gravitational Lagrangian is given by an arbitrary function of the Ricci Scalar R and of the trace T of the stress energy tensor $T_{ij}$. The field equations of this theory are derived from the Hilbert- Einstein type variational principle by taking the action.

$$S = \frac{1}{16\pi}\int [f(R,T) + L_m]\sqrt{-g}\, d^4x \qquad (1)$$

where $L_m$ is the matter Lagrangian density.

Stress energy tensor of matter is defined as

$$T_{ij} = \frac{-2}{\sqrt{-g}} \frac{\delta(\sqrt{-g}\, L_m)}{\delta^{ij}} \qquad (2)$$

and the trace by $T = g^{ij}T_{ij}$ respectively. By assuming that $L_m$ of matter depends only on the metric tensor components $g^{ij}$, we have obtain the field equations of $f(R,T)$ gravity as

$$f_R(R,T)R_{ij} - \frac{1}{2}f(R,T)g_{ij} + (g_{ij}\Box - \nabla_i\nabla_j)f_R(R,T) =$$

$$8\pi T_{ij} - f_T(R,T)T_{ij} - f_T(R,T)\theta_{ij} \qquad (3)$$

where $\theta_{ij} = -2T_{ij} + g_{ij}L_m - 2g^{lk}\frac{\partial^2 L_m}{\partial g^{ij}\partial g^{lm}}$ \qquad (4)

Here $f_R = \frac{\delta f(R,T)}{\delta R}$, $f_T = \frac{\delta f(R,T)}{\delta T}$, $\Box = \nabla^i\nabla_j$ and $\nabla^i$ is the covariant derivative. It may be noted that when $f(R,T) = f(R)\, Eq.\,(3)$ yields the field equations $f(R)$ gravity.

The problem of perfect fluids described an energy density $\rho$, pressure $p$ and four velocity $u^i$ is complicated since. There is no unique definition of the matter Lagrangian. However, here,

we assume that stress energy tensor of matter is given by

$$T_{ij} = (\rho + p)u_i u_j - p g_{ij} \qquad (5)$$

and the matter Lagrangian can be taken as $L_m = -p$ and we have

$$u^i \nabla_j u_i = 0, \quad u^i u_i = 1 \qquad (6)$$

Now with the use of Eq. (5) we obtain, for the variation of stress energy of perfect fluid, the expression

$$\theta_{ij} = -2T_{ij} - p g_{ij} \qquad (7)$$

Generally, the field equations also depend through the tensor $\theta_{ij}$, on the physical nature of the matter field. Hence in the case of $f(R,T)$ gravity, depending on the nature of the matter source, we obtain several theoretical models corresponding to each choice of $f(R,T)$. Assuming

$$f(R,T) = R + 2f(T) \qquad (8)$$

as a first choice where $f(T)$ is an arbitrary function of the trace of stress energy tensor of matter, we get the gravitational field equations of $f(R,T)$ gravity from Eq.(3) as( Harko et al.2011)

$$R_{ij} - \frac{1}{2} g_{ij} R = 8\pi T_{ij} - 2f'(T)T_{ij} - 2f'(T)\theta_{ij} + f(T)g_{ij} \qquad (9)$$

where the prime denotes differentiation with respect to the arguments. If the mater source is perfect fluid then the field equations of $f(R,T)$ gravity, in view of Eq.(7), become

$$R_{ij} - \frac{1}{2}g_{ij}R = 8\pi T_{ij} + 2f'(T)T_{ij} + [2pf'(T) + f(T)]g_{ij} \qquad (10)$$

## 3. Metric and field equations.

We consider the spatially homogeneous and anisotropic Bianchi Type-III metric in the form

$$ds^2 = dt^2 - A^2 dx^2 - e^{-2\beta x} B^2 dy^2 - C^2 dz^2 \qquad (11)$$

where A, B, C are functions of cosmic time t and β is a positive constant which can be taken as unity.

The energy momentum tensor for anisotropic dark energy is given by

$$T_i^j = diag\,[\rho, -\rho_x, -\rho_y, -\rho_z] \qquad (12)$$

$$= diag[\,1, -w_x, -w_y, -w_z\,]\,\rho$$

where ρ is the energy density of the fluid and $p_x, p_y, p_z$ are the pressures on the x, y and z-axis respectively. Here w is EoS parameter of the fluid with no deviation and $w_x$, $w_y$, $w_z$ are the EoS parameters in the directions of x, y and z axis respectively. The energy momentum tensor can be parameterized as

$$T_i^j = diag\,[1, -w, -(w+\gamma), -(w+\delta)]\rho \qquad (13)$$

For the sake of simplicity, we choose $w_x = w$ and the skewness parameters $\gamma$ and $\delta$ are the deviations from w on y and z axes respectively.

Now, with the choice of the function (Harko et al.2011)

$$f(T) = \lambda T \qquad (14)$$

where λ is a constant, the field equations of f(R,T) gravity given by Eq.(10) for the metric (11), can be written as

$$\frac{\dot{B}}{AB}+\frac{\ddot{A}\dot{C}}{AC}+\frac{\ddot{B}\dot{C}}{BC}-\frac{1}{A^2}=-\rho\left[(8\lambda+2\lambda)+\lambda(1-3w-\gamma-\delta)\right]-2\lambda p \qquad (15)$$

$$\frac{\dot{B}}{B}+\frac{\ddot{C}}{C}+\frac{\ddot{B}\dot{C}}{BC}=\rho\left[(8\lambda+2\lambda)w-\lambda(1-3w-\gamma-\delta)\right]-2\lambda p \qquad (16)$$

$$\frac{\ddot{A}}{A}+\frac{\ddot{C}}{C}+\frac{\ddot{A}\dot{C}}{AC}=\rho\left[(8\lambda+2\lambda)(w+\gamma)-\lambda(1-3w-\gamma-\delta)\right]-2\lambda p \qquad (17)$$

$$\frac{\ddot{A}}{A}+\frac{\ddot{B}}{B}+\frac{\ddot{A}\dot{B}}{AB}-\frac{1}{A^2}=\rho\left[(8\lambda+2\lambda)(w+\delta)-\lambda(1-3w-\gamma-\delta)\right]-2\lambda p \qquad (18)$$

$$\frac{\dot{A}}{A}-\frac{\dot{B}}{B}=0 \qquad (19)$$

where an overhead dot indicates differentiation with respect to t

The following are the parameters for the metric (11) which will be useful in solving the field equations:

The spatial volume is given by

$$V=a^3(t)=ABC \qquad (20)$$

where a(t) is the average scale factor of the universe.

The scalar expansion and shear scalar $\sigma^2$ are (kinematical parameters)

$$\theta = u^i_{;i} = \frac{\dot{A}}{A}+\frac{\dot{B}}{B}+\frac{\dot{C}}{C} \qquad (21)$$

$$\sigma^2 = \frac{1}{3}\left[\left(\frac{\dot{A}}{A}\right)^2+\left(\frac{\dot{B}}{B}\right)^2+\left(\frac{\dot{C}}{C}\right)^2-\frac{\dot{A}\dot{B}}{AB}-\frac{\dot{B}\dot{C}}{BC}-\frac{\dot{C}\dot{A}}{CA}\right] \qquad (22)$$

The average Hubble parameters and mean anisotropy parameter are defined as

$$H = \frac{1}{3}\theta = \left(\frac{\dot{A}}{A} + \frac{\dot{B}}{B} + \frac{\dot{C}}{C}\right) \tag{23}$$

$$\Delta = \frac{1}{3}\sum_{i=1}^{3}\left(\frac{H_i - H}{H}\right)^2 \tag{24}$$

## 4. Solutions and the model

Integration of Eq. (19) yields

$$A = k\,B \tag{25}$$

where k is a constant of integration which can be taken as unity without loss of any generality, so that we have

$$A = B \tag{26}$$

use of Eq. (26) reduces the field equations to the following form

$$\frac{\dot{B}^2}{B^2} + 2\frac{\ddot{B}\ddot{C}}{BC} - \frac{1}{B^2} = \rho\left[(8\bar{\lambda} + 2\lambda) + \lambda(1 - 3w - \gamma - \delta)\right] - 2\lambda p \tag{27}$$

$$\frac{\ddot{B}}{B} + \frac{\ddot{C}}{C} + \frac{\ddot{BC}}{BC} = \rho\left[(8\bar{\lambda} + 2\lambda)w - \lambda(1 - 3w - \gamma - \delta)\right] - 2\lambda p \tag{28}$$

$$\frac{\ddot{B}}{B} + \frac{\ddot{C}}{C} + \frac{\ddot{BC}}{BC} = \rho\left[(8\bar{\lambda} + 2\lambda)(w + \gamma) - \lambda(1 - 3w - \gamma - \delta)\right] - 2\lambda p \tag{29}$$

$$2\frac{\ddot{B}}{B} + \frac{\dot{B}^2}{B^2} - \frac{1}{B^2} = \rho\left[(8\bar{\lambda} + 2\lambda)(w + \delta) - \lambda(1 - 3w - \gamma - \delta)\right] - 2\lambda p \tag{30}$$

We have from Eqs. (28) and (29)

$$\gamma = 0 \tag{31}$$

Using Eq. (31) in Eqs. (27) – (30) we have

$$\frac{\ddot{B}}{B}+\frac{\ddot{C}}{C}-\frac{\dot{B}^2}{B^2}-\frac{\dot{B}\dot{C}}{BC}+\frac{1}{B^2}=\rho(8\lambda+2\lambda)(w+1) \quad (32)$$

$$\frac{2\ddot{B}}{B}-\frac{2\dot{B}\dot{C}}{BC}=\rho(8\lambda+2\lambda)(1+w+\delta) \quad (33)$$

Now, Eqs. (32) – (33) are two independent equations in five unknowns B, C, ρ, w and δ. Hence to find a determinate solution we need three more conditions.

(i) We use the time varying deceleration parameter q as suggested by Mishra et al .(2013, 2015)

$$q = \frac{-a\ddot{a}}{\dot{a}^2} = b(t) \quad (34)$$

which yield on integration (neglecting the integration constant) the scale factor

$$a(t) = (\sinh \alpha t)^{1/n} \quad (35)$$

(ii) We also use the fact that the scalar expansion θ is proportional to the shear scalar so that we have (Collins et al . 1980)

$$B = C^m \quad (36)$$

(iii) the EoS parameter w is proportional to the skewness parameter (mathematical condition) such that(Reddy et al.2013)

$$w + \delta = 0 \quad (37)$$

Now using Eqs. (21), (26), (35) and (36), we obtain the metric potentials of the model as

$$A = B = (\sinh \alpha t)^{\frac{3m}{n(2m+1)}}$$

(38)

$$C = (\sinh \alpha t)^{\frac{3}{n(2m+1)}}$$

where $\alpha$ and $n > 0$ are arbitrary constant using (38) the metric (11) can be written as

$$ds^2 = dt^2 - (\sinh \alpha t)^{\frac{6m}{n(2m+1)}}(dx^2 + e^{2x}dy^2) - (\sinh \alpha t)^{\frac{6}{n(2m+1)}}dz^2 \qquad (39)$$

## 5. Physical behavior of the model

To discuss the physical behavior of the model given by Eq. (39) we find the following physical and kinematical parameters of the model which are very important in the discussion of cosmology.

The spatial volume

$$V = (\sin h\alpha t)^{\frac{3}{n}} \qquad (40)$$

The average Hubble parameter is

$$H = \frac{3\alpha}{n} Cot\, h\alpha t \qquad (41)$$

The scalar expansion $\theta$ is

$$\theta = 3H = \frac{9\alpha}{n} Cot\, h\alpha t \qquad (42)$$

The deceleration parameter

$$q = \frac{d}{dt}\left(\frac{1}{H}\right) - 1 = \frac{n}{3}(1 - \tan h^2 \alpha t) - 1 \qquad (43)$$

The shear scalar is

$$\sigma^2 = \frac{3\alpha^2(m-1)^2}{n^2(2m+1)^2} \cot h^2(\alpha t) \tag{44}$$

The average anisotropy parameter is

$$\Delta = \frac{(6m^2+4m+2)}{(2m+1)^2} \tag{45}$$

Using Eqs (37) and (38) in Eq. (33) we obtain

$$\rho = -p = \frac{1}{((8x+2\lambda)n^2(2m+1)^2)} = \begin{bmatrix} 18m\alpha^2(m-1)\coth^2\alpha t \\ +6n\alpha^2(2m+1)(1-\cot h^2\alpha t) \end{bmatrix} \tag{46}$$

since we have in the case of accelerated expansion $\rho + p = 0$

Using Eqs. (32), (37), (38) and (46) we get

w = - δ =

-1 + $9\alpha^2$ (1-m) cot h²αt + $3m\alpha^2$(2m+1)(1+m)(1-coth²αt)

$$\frac{-n^2(2m+1)^2(\sin h^2\alpha t)^{\frac{-6m}{n(2m+1)}}}{18m\alpha^2(m-1)ath^2\alpha t + bn\alpha^2(2m+1)(1-\coth^2\alpha t)} \tag{47}$$

Thus Eq. (39) represents Bianchi type-III anisotropic dark energy universe with the above cosmological parameters. It may be observed that the model is free from initial singularity. The spatial volume of the universe increases as t increases. It can also be seen that H, θ, σ, ρ, δ, w and p are functions of t and vanish for large t while they all diverge for t = 0. Also, since $\frac{\sigma^2}{\theta^2} \neq 0$, the model does not approach isotropy for large t. However, the model becomes isotropic and shear free because in this case the model reduces to

$$ds^2 = dt^2 - A^2(t)[dx^2 - e^{-2x}dy^2 - dz^2] \tag{47}$$

## 6. Conclusions

In this paper, we have investigated anisotropic Bianchi type-III dark energy model in f (R, T) gravity proposed by Harko et al. (2011). The gravitational field equations have been solved using a variable deceleration parameter suggested by Mishra et al (2013). We have also used the fact that the scalar expansion is proportional to the shear scalar. The model obtained represents Bianchi type-III dark energy universe in f(R, T) gravity. The cosmological parameters of the model have been computed and the physical behavior of the universe is discussed.

## Compliance with ethical standards

The authors declare that they have no potential conflict and will abide by the ethical standards of this journal .